# Unified Theory of Wave-Particle Duality, the Schrödinger Equations, and Quantum Diffraction

**Greyson Gilson**

Mulith Inc.
30 Chestnut Street # 32
Nashua, New Hampshire 03060-3361, USA
Email: greyson.gilson@mulithinc.com
4 September 2014

## ABSTRACT

Individual quantum objects display inseparable coexisting wave-like properties and particle-like properties; such inseparable coexistence can seem paradoxical and mind-boggling. The apparent paradox is resolved by the unified theory of wave-particle duality developed in this paper. Based on the unified theory of wave-particle duality, a straightforward derivation of the Schrödinger equations is presented where previously no such derivation was considered to be possible. A new theory of quantum diffraction is subsequently developed.

**PLAN OF THE PAPER**

## I. INTRODUCTION

Quantum objects are inseparably associated with wave-like properties and particle-like properties (Refer to Appendix A); this twofold character is known as wave-particle duality. The name (electron, photon, atom, molecule, etc.) linked to its particle-like properties is used to identify a quantum object. Several wave-particle duality demonstrations have been recorded as movies; perhaps the most well-known of these movies has been produced by Hitachi, Ltd. and attributed to Akira Tonomura[1].

A quantum object's wave-like properties can seem to be associated with a very large region of physical space; a quantum object's particle-like properties can seem to be associated with a very small region of physical space. A single entity that seems to be inseparably associated with both a very large region of physical space and a very small region of physical space can be regarded as paradoxical. Neither the wave viewpoint nor the particle viewpoint is correct[2]. A unified theory of wave-particle duality whereby the apparent paradox is resolved is developed in this paper.

After developing the unified theory of wave-particle duality, a straightforward treatment that leads inexorably to the Schrödinger equations is presented. Subjective plausibility arguments that vary from author to author are ordinarily used to introduce the Schrödinger equations (Refer to Appendix B). No derivation is attempted by these authors; apparently, derivation of the Schrödinger equations is thought to be impossible. Rather, the Schrödinger equations are pragmatically justified by the many successes that have been attained by assuming they are correct. The Schrödinger equations provide the basis for a correct analysis of all kinds of molecular, atomic and nuclear systems[3]. The Schrödinger equations are shown to be relativistically invariant (as clarified by Barut[4], relativistic invariance does not necessarily imply covariance) and to constitute laws of physics.

The unified theory of wave-particle duality provides a solid basis for deriving the Schrödinger equations. Consequently, the unified theory of wave-particle duality shares the success demonstrated by the Schrödinger equations.

A fundamental treatment of quantum diffraction that is independent of the Schrödinger equations follows the derivation of the Schrödinger equations. Quantum diffraction occurs, as a consequence of their wave properties, when quantum objects pass through an aperture or pass the edge of an obstacle; accordingly, quantum objects can spread into regions that are not directly exposed to them.

## II. UNIFIED THEORY OF WAVE-PARTICLE DUALITY

### WAVE-LIKE PROPERTIES

As a principal ingredient, the unified theory of wave-particle duality includes the hypothesis that the optical differential wave equation is applicable to all quantum objects. Thus, the quantum amplitude $\Psi$ linked to a quantum object satisfies the differential wave equation

$$\nabla^2 \Psi = \frac{\partial^2 \Psi}{\partial t^2} \tag{2.1}$$

In this equation, $t$ is time, $\nabla^2$ is the Laplacian operator, and the speed of wave propagation (unity – the speed of light) is the same for all quantum objects; the units used for time are the same as those used for distance. $\Psi$ represents a space-filling and time-dependent physical field.

Solutions of equation (2.1) that have the form

$$\Psi(\mathbf{r},t) = \psi(\mathbf{r})\varphi(t) \tag{2.2}$$

where $\mathbf{r}$ is the position vector for a point in space, can be obtained by separation of variables. After substituting equation (2.2) into equation (2.1) and then dividing both sides of the result by $\psi(\mathbf{r})\varphi(t)$

$$\frac{\nabla^2 \psi(\mathbf{r})}{\psi(\mathbf{r})} = \frac{1}{\varphi(t)} \frac{\partial^2 \varphi(t)}{\partial t^2} \tag{2.3}$$

follows.

The variables on the left hand side of equation (2.3) are independent of the variable on the right hand side of equation (2.3). Consequently, each side of the equation is necessarily equal to the same constant. Accordingly

$$\frac{\nabla^2 \psi(\mathbf{r})}{\psi(\mathbf{r})} = -k^2 \tag{2.4}$$

and

$$\frac{1}{\varphi(t)} \frac{\partial^2 \varphi(t)}{\partial t^2} = -k^2 \tag{2.5}$$

where $-k^2$ is the separation constant. Equations (2.4) and (2.5) can be written as

$$\left(\nabla^2 + k^2\right)\psi\left(\mathbf{r}\right) = 0 \tag{2.6}$$

which is the Helmholtz equation and

$$\left(\frac{\partial^2}{\partial t^2} + k^2\right)\varphi\left(t\right) = 0 \tag{2.7}$$

respectively. The spatial dependence of $\Psi\left(\mathbf{r}, t\right)$ satisfies equation (2.6) while the time dependence of $\Psi\left(\mathbf{r}, t\right)$ satisfies equation (2.7).

**TIME DEPENDENCE**

Equation (2.7) is a partial differential equation that involves only one variable and can consequently be written as

$$\frac{d^2\varphi}{dt^2} + k^2\varphi = 0 \tag{2.8}$$

which is an ordinary differential equation. Equation (2.8) yields the solution

$$\varphi\left(t\right) = C_+ \exp\left(-ikt\right) + C_- \exp\left(ikt\right) \tag{2.9}$$

where $C_+$ and $C_-$ are constants of integration.

Oscillations of temporal frequency

$$\nu = \frac{k}{2\pi} \tag{2.10}$$

where $\nu$ and $k$ are positive real numbers, are described by equation (2.9). As observed at a particular point in space the quantum amplitude oscillates at the frequency $\nu$ and is related by the equation

$$\lambda\,\nu = 1 \tag{2.11}$$

to the wavelength $\lambda$ (a positive real number) that is linked to the quantum object. Equation (2.11) is a fundamental equation of wave motion.

The wave number

$$k = \frac{2\pi}{\lambda} \tag{2.12}$$

can be obtained after equation (2.11) has been substituted into equation (2.10). Furthermore

$$\varphi(t) = C_+ \exp(-i2\pi\nu t) + C_- \exp(i2\pi\nu t) \tag{2.13}$$

results when equation (2.10) is substituted into equation (2.9).

**WAVE PROPAGATION**

Equation (2.2) can be written, in any three-dimensional rectangular coordinate system, as

$$\Psi(x, y, z, t) = \psi(z, y, z)\varphi(t) \tag{2.14}$$

where

$$\Psi(x, y, z, t) = \Psi(\mathbf{r}, t) \tag{2.15}$$

and

$$\psi(x, y, z) = \psi(\mathbf{r}) \tag{2.16}$$

have been introduced. Substitution of equation (2.13) into equation (2.14) yields

$$\Psi(x, y, z, t) = \psi(x, y, z)\left[C_+ \exp(-i2\pi\nu t) + C_- \exp(i2\pi\nu t)\right] \tag{2.17}$$

directly.

Equation (2.6), the Helmholtz equation, is given by

$$\left(\frac{\partial^2}{\partial x^2} + \frac{\partial^2}{\partial y^2} + \frac{\partial^2}{\partial z^2} + k^2\right)\psi(x, y, z) = 0 \tag{2.18}$$

in the three-dimensional rectangular coordinate system. In this equation the Laplacian operator

$$\nabla^2 = \frac{\partial^2}{\partial x^2} + \frac{\partial^2}{\partial y^2} + \frac{\partial^2}{\partial z^2} \tag{2.19}$$

has been expressed in rectangular coordinates.

The three-dimensional Fourier transform of $\psi(x, y, z)$

$$\tilde{\psi}(\nu_x, \nu_y, \nu_z) = \int_{-\infty}^{+\infty} \int_{-\infty}^{+\infty} \int_{-\infty}^{+\infty} \psi(x, y, z) \exp\left[-i2\pi(\nu_x x + \nu_y y + \nu_z z)\right] dz\, dy\, dx \qquad (2.20)$$

can now be conveniently introduced. Here, the reciprocal variables $\nu_x$, $\nu_y$ and $\nu_z$, commonly known as spatial frequencies, are needed to define the three-dimensional Fourier transform of $\psi(x, y, z)$. The corresponding three-dimensional inverse Fourier transform, given by

$$\psi(x, y, z) = \int_{-\infty}^{+\infty} \int_{-\infty}^{+\infty} \int_{-\infty}^{+\infty} \tilde{\psi}(\nu_x, \nu_y, \nu_z) \exp\left[i2\pi(x\nu_x + y\nu_y + z\nu_z)\right] d\nu_z\, d\nu_y\, d\nu_x \qquad (2.21)$$

can also be conveniently introduced. The Fourier transform pair expresses $\tilde{\psi}(\nu_x, \nu_y, \nu_z)$ and $\psi(x, y, z)$ as linear combinations of three-dimensional complex exponential functions.

The exponent in the integrand of equation (2.21) can be written as

$$i2\pi(x\nu_x + y\nu_y + z\nu_z) = i\left(\frac{2\pi}{\lambda}\right)\left(x\cos\theta_x + y\cos\theta_y + z\cos\theta_z\right) \qquad (2.22)$$

where $\theta_x$, $\theta_y$ and $\theta_z$ are the angles between the directions of wave propagation and the $x$-, $y$- and $z$-axes, respectively. In addition, the direction cosines

$$\begin{pmatrix} \cos\theta_x \\ \cos\theta_y \\ \cos\theta_z \end{pmatrix} = \begin{pmatrix} \lambda\nu_x \\ \lambda\nu_y \\ \lambda\nu_z \end{pmatrix} \qquad (2.23)$$

have been introduced. Explicit expressions for the spatial frequencies $\nu_x$, $\nu_y$ and $\nu_z$ are provided by

$$\begin{pmatrix} \nu_x \\ \nu_y \\ \nu_z \end{pmatrix} = \begin{pmatrix} \nu\cos\theta_x \\ \nu\cos\theta_y \\ \nu\cos\theta_z \end{pmatrix} \qquad (2.24)$$

where equation (2.11) has been recalled. Equation (2.24) is the column vector representation of the temporal frequency vector $\nu$.

The inner product of the temporal frequency vector with itself is given by

$$\nu_x{}^2 + \nu_y{}^2 + \nu_z{}^2 = \nu^2 \left( \cos^2 \theta_x + \cos^2 \theta_y + \cos^2 \theta_z \right) \tag{2.25}$$

where

$$\cos^2 \theta_x + \cos^2 \theta_y + \cos^2 \theta_z = 1 \tag{2.26}$$

constitutes a well-known fundamental property of direction cosines. After using some trigonometry,

$$\sin^2 \theta_z = \cos^2 \theta_x + \cos^2 \theta_y \tag{2.27}$$

which is equivalent to

$$\sin \theta_z = \pm \sqrt{\cos^2 \theta_x + \cos^2 \theta_y} \tag{2.28}$$

follows from equation (2.26) readily.

Substitution of equation (2.26) into equation (2.25) leads to

$$\nu_x{}^2 + \nu_y{}^2 + \nu_z{}^2 = \nu^2 \tag{2.29}$$

easily. In turn, substitution of equation (2.11) into equation (2.29) yields

$$\nu_x{}^2 + \nu_y{}^2 + \nu_z{}^2 = \frac{1}{\lambda^2} \tag{2.30}$$

a result that will be used later.

Substitution of equation (2.21) into equation (2.17) leads to

$$\begin{aligned}
\Psi\left( x, y, z, t \right) = {}& \\
& C_+ \int_{-\infty}^{+\infty} \int_{-\infty}^{+\infty} \int_{-\infty}^{+\infty} \tilde{\psi}\left( \nu_x, \nu_y, \nu_z \right) \exp\left[ i 2\pi \left( x\nu_x + y\nu_y + z\nu_z - \nu t \right) \right] d\nu_z \, d\nu_y \, d\nu_x \\
& + C_- \int_{-\infty}^{+\infty} \int_{-\infty}^{+\infty} \int_{-\infty}^{+\infty} \tilde{\psi}\left( \nu_x, \nu_y, \nu_z \right) \exp\left[ i 2\pi \left( x\nu_x + y\nu_y + z\nu_z + \nu t \right) \right] d\nu_z \, d\nu_y \, d\nu_x
\end{aligned} \tag{2.31}$$

The two terms in equation (2.31) describe two superpositions of plane waves that propagate in opposite directions.

**PARTICLE-LIKE PROPERTIES**

Each individual quantum object is assumed to be associated with a mass $m$, energy $E$ and momentum $\mathbf{p}$ such that the momenergy relationships of special relativity[5] are satisfied. Thus, the square of the magnitude of the momenergy 4-vector associated with a quantum object is given by

$$E^2 - p^2 = m^2 \qquad (2.32)$$

where

$$p = |\mathbf{p}| \qquad (2.33)$$

is the magnitude of $\mathbf{p}$. No explicit location is assumed to be associated with a quantum object.

**COMBINED WAVE-LIKE AND PARTICLE-LIKE PROPERTIES**

The well-known Planck-Einstein relation[6]

$$E = h\nu \qquad (2.34)$$

where $h$ is Planck's constant, can now be introduced. Substitution of equation (2.10) into equation (2.34) leads to

$$E = \hbar k \qquad (2.35)$$

where the reduced Planck constant

$$\hbar = \frac{h}{2\pi} \qquad (2.36)$$

has been introduced. The relationship

$$k = \frac{E}{\hbar} \qquad (2.37)$$

can be obtained from equation (2.35) trivially.

Substitution of equation (2.32) into equation (2.37) leads to

$$k = \frac{\sqrt{m^2 + p^2}}{\hbar} \qquad (2.38)$$

easily. The positive square root has been chosen because $k$ is a positive real number. Subsequent substitution of equation (2.12) into equation (2.38) yields

$$\lambda = \frac{h}{\sqrt{m^2 + p^2}} \tag{2.39}$$

where equation (2.36) has been invoked.

The treatment thus far is applicable to massless quantum objects (where $m = 0$) and to massive quantum objects (where $m \neq 0$) alike.

**DEBROGLIE THEORY**

Association of wave-like behavior with quantum objects began with the work of Louis deBroglie (1924) and his postulate that any quantum object with momentum $p$ is linked to a wavelength $\lambda_d$ such that

$$\lambda_d = \frac{h}{p} \tag{2.40}$$

where $\lambda_d$ is known as the deBroglie wavelength associated with the quantum object. This wavelength was previously known to be associated with massless photons; deBroglie postulated that it is also associated with massive quantum objects.

Equation (2.39) reduces to equation (2.40) for the special case where $m = 0$. However, this special case does not include massive quantum objects. Thus, equation (2.40) is applicable to massless quantum objects (where $m = 0$) but is not applicable to quantum objects with mass (where $m \neq 0$) as deBroglie intended. Equation (2.40) is not an element of fundamental physics. Nevertheless, the deBroglie wavelength is very often accepted (erroneously) as an aspect of fundamental physics.

DeBroglie associated the location of a quantum object with a localized pulse composed of a superposition of waves moving at various speeds. The quantum object's speed was identified as the group velocity (less than the speed of light) of the superposition of waves.

In accord with equation (2.1), only one wave is associated with a quantum object. The notion of associating a superposition of waves and a group velocity with a quantum object is at variance with the unified theory of wave-particle duality.

Most early and current understanding of wave-particle duality has its roots in the work of deBroglie. A good concise and modern review of Louis deBroglie's theory and subsequent developments is provided by A. P. French and Edwin F. Taylor[7].

## III. SCHRÖDINGER EQUATIONS

### ENERGY

Classical energy is given by

$$E_c = \frac{p^2}{2m} + V(\mathbf{r}) + V_0 \tag{3.1}$$

where the spatially dependent potential energy $V(\mathbf{r})$ and the constant potential energy $V_0$ (which can be chosen at will) have been introduced. The relationship

$$\frac{p^2}{2m} = \frac{E^2}{2m} - \frac{m}{2} \tag{3.2}$$

can be obtained by dividing both sides of equation (2.32) by $2m$ and rearranging the result. After substituting equation (3.2) into equation (3.1) and introducing

$$V_0 = \frac{m}{2} + E_c - E \tag{3.3}$$

the relationship

$$\frac{E^2}{2m} = E - V(\mathbf{r}) \tag{3.4}$$

can be found easily.

### TIME-INDEPENDENT SCHRÖDINGER EQUATION

Substitution of equation (2.37) into equation (2.6) leads to

$$\nabla^2 \psi(\mathbf{r}) = -\left(\frac{E}{\hbar}\right)^2 \psi(\mathbf{r}) \tag{3.5}$$

readily. Equation (3.5) can be rearranged and written as

$$-\frac{\hbar^2}{2m}\nabla^2\psi(\mathbf{r})=\left(\frac{E^2}{2m}\right)\psi(\mathbf{r}) \qquad (3.6)$$

upon division by $2m$. Substitution of equation (3.4) into equation (3.6) leads to

$$-\frac{\hbar^2}{2m}\nabla^2\psi(\mathbf{r})+V(\mathbf{r})\psi(\mathbf{r})=E\psi(\mathbf{r}) \qquad (3.7)$$

without fanfare. Equation (3.7) is the time-independent Schrödinger equation.

**TIME-DEPENDENT SCHRÖDINGER EQUATION**

After multiplying both sides by $\varphi(t)$

$$-\frac{\hbar^2}{2m}\nabla^2\psi(\mathbf{r})\varphi(t)+V(\mathbf{r})\psi(\mathbf{r})\varphi(t)=E\psi(\mathbf{r})\varphi(t) \qquad (3.8)$$

can be obtained from equation (3.7). Equation (3.8) can be written as

$$-\frac{\hbar^2}{2m}\nabla^2\Psi(\mathbf{r},t)+V(\mathbf{r})\Psi(\mathbf{r},t)=E\Psi(\mathbf{r},t) \qquad (3.9)$$

where equation (2.2) has been invoked.

Substitution of equations (2.34) and (2.36) into equation (2.13) leads to

$$\varphi(t)=C_+\exp\left[-i\left(\frac{E}{\hbar}\right)t\right]+C_-\exp\left[i\left(\frac{E}{\hbar}\right)t\right] \qquad (3.10)$$

Equation (3.10) reduces to

$$\varphi(t)=C_+\exp\left[-i\left(\frac{E}{\hbar}\right)t\right] \qquad (3.11)$$

where, for the present special case

$$C_-=0 \qquad (3.12)$$

has been chosen. Substitution of equation (3.11) into equation (2.2) yields

$$\Psi\left(\mathbf{r},t\right)=C_{+}\psi\left(\mathbf{r}\right)\exp\left[-i\left(\frac{E}{\hbar}\right)t\right] \tag{3.13}$$

Equation (3.13) can be differentiated with respect to $t$ and rearranged to obtain

$$\Psi\left(\mathbf{r},t\right)=i\left(\frac{\hbar}{E}\right)\frac{\partial\Psi\left(\mathbf{r},t\right)}{\partial t} \tag{3.14}$$

Substitution of equation (3.14) into the right hand side of equation (3.9) yields

$$-\frac{\hbar^{2}}{2m}\nabla^{2}\Psi\left(\mathbf{r},t\right)+V\left(\mathbf{r}\right)\Psi\left(\mathbf{r},t\right)=i\hbar\frac{\partial\Psi\left(\mathbf{r},t\right)}{\partial t} \tag{3.15}$$

Equation (3.15) is the time-dependent Schrödinger equation.

**AUXILIARY TIME-DEPENDENT SCHRÖDINGER EQUATION**

Equation (3.10) reduces to

$$\varphi\left(t\right)=C_{-}\exp\left[i\left(\frac{E}{\hbar}\right)t\right] \tag{3.16}$$

where, for the present special case

$$C_{+}=0 \tag{3.17}$$

has been chosen. Substitution of equation (3.16) into equation (2.2) yields

$$\Psi\left(\mathbf{r},t\right)=C_{-}\psi\left(\mathbf{r}\right)\exp\left[i\left(\frac{E}{\hbar}\right)t\right] \tag{3.18}$$

Equation (3.18) can be differentiated with respect to $t$ and rearranged to obtain

$$\Psi\left(\mathbf{r},t\right)=-i\left(\frac{\hbar}{E}\right)\frac{\partial\Psi\left(\mathbf{r},t\right)}{\partial t} \tag{3.19}$$

Substitution of equation (3.19) into the right hand side of equation (3.9) yields

$$-\frac{\hbar^{2}}{2m}\nabla^{2}\Psi\left(\mathbf{r},t\right)+V\left(\mathbf{r}\right)\Psi\left(\mathbf{r},t\right)=-i\hbar\frac{\partial\Psi\left(\mathbf{r},t\right)}{\partial t} \tag{3.20}$$

Equation (3.20) is the auxiliary time-dependent Schrödinger equation.

## SCHRÖDINGER EQUATIONS AS LAWS OF PHYSICS

The unified theory of wave-particle duality has been used to derive the Schrödinger equations. The Schrödinger equations are generally accepted, by postulate rather than derivation, to be laws of physics.

The Schrödinger equations provide a basis for analyzing many kinds of systems (molecular, atomic, and nuclear) in a particular inertial reference frame. The success of the Schrödinger equations constitutes a basis for accepting them, their derivations, and the unified theory of wave-particle duality which makes such derivations possible. This acceptance is completely justified in the favored inertial reference frame.

In accord with the principle of relativity, all physical laws must be the same in all inertial reference frames, i.e., all physical laws must be Lorentz invariant. Equation (2.1) is Lorentz invariant[8] and reduces, by means of the procedure presented in this paper, to the Schrödinger equations. As a result, equation (2.1) constitutes a unique Lorentz invariant form of the Schrödinger equations[9]. Consequently, the Schrödinger equations are relativistically invariant.

Conservation of the number of massive quantum objects present occurs when the Schrödinger equations are applicable[10]. This happens when positrons are absent, nuclei are stable, and energy transfers lie below the threshold for electron-positron pair production. These low-energy conditions are often considered to be *non-relativistic*. Correspondingly, the term *relativistic* is often restricted for use when the energies involved are high enough to permit non-conservation of the number of massive quantum objects present. This use of terminology does not change the fact that the Schrödinger equations are relativistically invariant.

Relativistic invariance does not imply covariance[11]. Relativistic invariance of the Schrödinger equations is based on reduction from a covariant law. Because of their relativistic invariance and their success in analyzing many kinds of systems in particular inertial reference frames, the Schrödinger equations constitute laws of physics. These laws of physics are applicable when energies are low enough to assure that the number of massive quantum objects present is conserved.

In accord with the unified theory of wave-particle duality, quantum objects are linked to particle-like properties, but not to particles. This is consistent with Blood's[12] finding that there is no evidence for particles and with Hobson's[13] finding that there are no particles, only fields.

The Schrödinger equations are field equations, not particle equations. Rather than describing particle motion, the time-dependent Schrödinger equation describes a time-dependent field $\Psi(\mathbf{r}, t)$ throughout a spatial region. Following Hobson, the Schrödinger field is a space-filling physical field whose value at any spatial point is the probability amplitude for an interaction to occur within an infinitesimal region surrounding the point.

**RELATIVE INTERACTION PROBABILITY DENSITY**

As indicated previously, the Schrödinger field $\Psi(\mathbf{r},t)$ is a space-filling physical field whose value at any spatial point is the probability amplitude for an interaction to occur within an infinitesimal region surrounding the point. The relative interaction probability density

$$P(\mathbf{r}) = \psi^*(\mathbf{r})\psi(\mathbf{r}) \tag{3.21}$$

where the superscript asterisk denotes complex conjugation, can be introduced to specify the relative probability for an interaction to occur within an infinitesimal region surrounding the point $\mathbf{r}$. Equation (3.21) can be written as

$$P(x,y,z) = \psi^*(x,y,z)\psi(x,y,z) \tag{3.22}$$

after equation (2.16) has been recalled. The relative interaction probability density

$$P(\mathbf{r}) = P(x,y,z) \tag{3.23}$$

specifies the relative probability that a quantum object can interact within an infinitesimal region surrounding the point $\mathbf{r} = (x,y,z)$.

# IV. QUANTUM DIFFRACTION

**FUNDAMENTALS**

As a consequence of their wave-like properties quantum objects can exhibit diffraction phenomena. When waves pass through an aperture or pass the edge of an obstacle they can spread into regions that are not directly exposed to them. This phenomenon is called diffraction and is a property of all kinds of propagating waves.

Quantum diffraction occurs when the lateral extent of propagating waves linked to quantum objects is spatially limited. After passing the spatial constraints the waves are diffracted and propagate toward various surfaces where they form diffraction patterns. The quantum objects that form a diffraction pattern define a relative interaction probability density on the surfaces where the diffraction patterns form.

Diffraction patterns are formed by the apparently random arrival of one quantum object at a time[14, 15]. A diffraction pattern that is formed in a sufficiently feeble manner shows no evidence of having wave-like properties. Rather, diffraction patterns are built up by individual quantum objects that act independently. An inherent granularity exists during the early portion of the diffraction pattern build-up.

A very large number of quantum objects contribute when diffraction pattern formation occurs. The granularity linked to individual quantum objects vanishes when the diffraction pattern is completely formed.

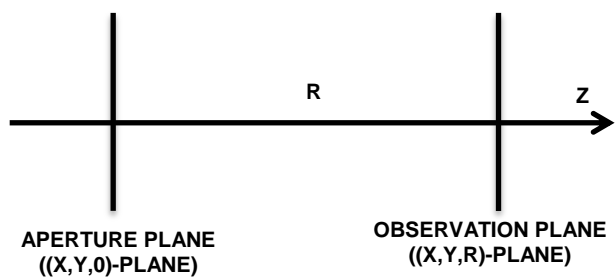

Figure 1. Diffraction apparatus.

A diffraction pattern that is formed in a sufficiently strong manner shows no evidence of being built up by individual quantum objects. Rather, the entire diffraction pattern appears to be formed as a single occurrence.

During the diffraction process quantum objects propagate away from regions of lateral confinement and ultimately arrive at an observation surface. A simple diffraction apparatus is illustrated in Figure 1. In this apparatus regions of a wave's lateral confinement are idealized as existing in an aperture plane; similarly, the observation surface is idealized as an observation plane. The observation plane is parallel to the aperture plane. Although generalizations to non-planar surfaces can be made, treatment of them is beyond the scope of this paper.

Initially, a quantum amplitude that is incident perpendicular to the aperture plane is transmitted through apertures in the aperture plane or reflected from physical objects in the aperture plane. Subsequently, the resulting quantum amplitude is distributed in a definite configuration on the side of the aperture plane nearest to the observation plane. A portion of this quantum amplitude propagates from the aperture plane to the observation plane.

The portion of the quantum amplitude that arrives at the observation plane forms a relative interaction probability density on the observation plane. The quantum objects that arrive at the observation surface are distributed in accord with the relative interaction probability density and form a diffraction pattern. The diffraction pattern may or may not be observed.

Referring to Figure 1, a right-handed rectangular coordinate system can be conveniently introduced to facilitate treating quantum diffraction mathematically. In this coordinate system, the aperture plane is the $(x, y, 0)$-plane and the observation plane is the $(x, y, R)$-plane; the observation plane is separated from the aperture plane by the distance $R$. The $z$-axis intersects the $(x, y, 0)$-plane at the origin of coordinates. The positive $z$-direction is the direction from the aperture plane toward the observation plane. The portion of the quantum amplitude that propagates away from the aperture plane toward the observation plane travels in the positive $z$-direction.

**INCIDENT WAVES**

Equation (2.31) describes two superpositions of plane waves that propagate in opposite directions. For waves that that propagate in the positive $z$-direction

$$C_- = 0 \tag{4.1}$$

is required.

As two results, equations (2.13) and (2.31) reduce to

$$\varphi(t) = C_+ \exp(-i2\pi\nu t) \tag{4.2}$$

and

$$\Psi(x, y, z, t) = C_+ \int_{-\infty}^{+\infty} \int_{-\infty}^{+\infty} \int_{-\infty}^{+\infty} \tilde{\psi}(\nu_x, \nu, \nu_z) \exp\left[i2\pi\left(x\nu_x + y\nu_y + z\nu_z - \nu t\right)\right] d\nu_z \, d\nu_y \, d\nu_x \tag{4.3}$$

respectively. The plane waves described by equation (4.3) propagate in the positive $z$-direction.

For waves that propagate in the positive z-direction

$$\nu_z > 0 \tag{4.4}$$

is required. Equation (2.29) can now be solved to obtain

$$\nu_z = \sqrt{\nu^2 - \nu_x{}^2 - \nu_y{}^2} \tag{4.5}$$

where inequality (4.4) has been used to support rejecting the negative square root. In addition, inequality (4.4) can be substituted into the third row of equation (2.23) to obtain

$$\cos\theta_z > 0 \tag{4.6}$$

because $\lambda$, given by equation (2.39), is a positive real number.

Waves described by equation (4.3) interact with apertures or physical features in the aperture plane. As a result of such interaction, a definite configuration of quantum objects forms on any arbitrary plane between the aperture plane and the observation plane. This configuration of quantum objects is linked to a quantum amplitude.

**ANGULAR SPECTRUM**

Let $\psi\left(x,y,z\right)$ be the quantum amplitude of quantum objects (or, more simply, the quantum amplitude) at an arbitrary point between the aperture plane and the observation plane. Then

$$\tilde{\psi}\left(v_x,v_y;z\right)=\int_{-\infty}^{+\infty}\int_{-\infty}^{+\infty}\psi\left(x,y,z\right)\exp\left[-i2\pi\left(v_x x+v_y y\right)\right]dy\,dx \qquad (4.7)$$

is the two-dimensional Fourier transform of $\psi\left(x,y,z\right)$ on the plane defined by an arbitrary constant value of $z$;

$$\psi\left(x,y,z\right)=\int_{-\infty}^{+\infty}\int_{-\infty}^{+\infty}\tilde{\psi}\left(v_x,v_y;z\right)\exp\left[i2\pi\left(xv_x+yv_y\right)\right]dv_y\,dv_x \qquad (4.8)$$

is the corresponding two-dimensional inverse Fourier transform. The function $\tilde{\psi}\left(v_x,v_y;z\right)$ is known as the angular spectrum of quantum objects (or, more simply, the angular spectrum) on the plane defined by an arbitrary constant value of $z$.

The relative spatial frequency interaction probability density given by

$$\tilde{P}\left(v_x,v_y;z\right)=\tilde{\psi}^{*}\left(v_x,v_y;z\right)\tilde{\psi}\left(v_x,v_y;z\right) \qquad (4.9)$$

can be conveniently introduced. The relative spatial frequency interaction probability density specifies the relative probability that a quantum object with spatial frequencies that are infinitesimally near $v_x$ and $v_y$ can interact within an infinitesimal distance near the $z$-plane

**PROPAGATION OF THE ANGULAR SPECTRUM**

The two-dimensional Fourier transform pairs defined by equations (4.7) and (4.8) can be used to treat propagation of the angular spectrum. The method used here is an extension of a method Goodman[16,17] has used previously.

Substitution of equation (4.8) into equation (2.18) yields

$$\left(\frac{\partial^2}{\partial x^2}+\frac{\partial^2}{\partial y^2}+\frac{\partial^2}{\partial z^2}+k^2\right)\int_{-\infty}^{+\infty}\int_{-\infty}^{+\infty}\tilde{\psi}\left(v_x,v_y;z\right)\exp\left[i2\pi\left(xv_x+yv_y\right)\right]dv_y\,dv_x=0 \qquad (4.10)$$

which can be written as

$$\int_{-\infty}^{+\infty}\int_{-\infty}^{+\infty}\left(\frac{\partial^2}{\partial x^2}+\frac{\partial^2}{\partial y^2}+\frac{\partial^2}{\partial z^2}+k^2\right)\tilde{\psi}\left(\nu_x,\nu_y;z\right)\exp\left[i2\pi\left(x\nu_x+y\nu_y\right)\right]d\nu_y\,d\nu_x=0 \qquad (4.11)$$

because the integration variables differ from the differentiation variables. Equation (4.11) reduces to

$$\int_{-\infty}^{+\infty}\int_{-\infty}^{+\infty}\left\{\frac{\partial^2}{\partial z^2}\left[\tilde{\psi}\left(\nu_x,\nu_y;z\right)\right]+\left(k^2-4\pi^2\nu_x^{\;2}-4\pi^2\nu_y^{\;2}\right)\tilde{\psi}\left(\nu_x,\nu_y;z\right)\right\}$$
$$X\,\exp\left[i2\pi\left(x\nu_x+y\nu_y\right)\right]d\nu_y\,d\nu_x=0 \qquad (4.12)$$

or, equivalently,

$$\int_{-\infty}^{+\infty}\int_{-\infty}^{+\infty}\left\{\frac{\partial^2}{\partial z^2}\left[\tilde{\psi}\left(\nu_x,\nu_y;z\right)\right]+4\pi^2\nu_z^{\;2}\,\tilde{\psi}\left(\nu_x,\nu_y;z\right)\right\}\exp\left[i2\pi\left(x\nu_x+y\nu_y\right)\right]d\nu_y\,d\nu_x=0 \quad (4.13)$$

after equations (2.12), (2.29) and (2.30) are invoked.

Each component function in the integrand of equation (4.13) satisfies the equation independently and the partial differential equation

$$\frac{\partial^2}{\partial z^2}\left[\tilde{\psi}\left(\nu_x,\nu_y;z\right)\right]+4\pi^2\nu_z^{\;2}\,\tilde{\psi}\left(\nu_x,\nu_y;z\right)=0 \qquad (4.14)$$

necessarily follows. Equation (4.14) involves only one independent variable and can therefore be written as

$$\frac{d^2}{dz^2}\left[\tilde{\psi}\left(\nu_x,\nu_y;z\right)\right]+4\pi^2\nu_z^{\;2}\,\tilde{\psi}\left(\nu_x,\nu_y;z\right)=0 \qquad (4.15)$$

which is an ordinary differential equation. Equation (4.15) can be solved to yield the solution

$$\tilde{\psi}\left(\nu_x,\nu_y;z\right)=D_+\exp\left(-i2\pi z\nu_z\right)+D_-\exp\left(i2\pi z\nu_z\right) \qquad (4.16)$$

where $D_+$ and $D_-$ are constants of integration.

Substitution of equation (4.16) into equation (4.8) yields

$$\psi\left(x,y,z\right) = D_+\int_{-\infty}^{+\infty}\int_{-\infty}^{+\infty}\exp\left[i2\pi\left(xv_x + yv_y - zv_z\right)\right]dv_y\,dv_x$$
$$+ D_-\int_{-\infty}^{+\infty}\int_{-\infty}^{+\infty}\exp\left[i2\pi\left(xv_x + yv_y + zv_z\right)\right]dv_y\,dv_x$$

(4.17)

readily. Furthermore

$$\Psi\left(x,y,z,t\right) = C_+D_+\int_{-\infty}^{+\infty}\int_{-\infty}^{+\infty}\exp\left[i2\pi\left(xv_x + yv_y - zv_z - vt\right)\right]dv_y\,dv_x$$
$$+ C_+D_-\int_{-\infty}^{+\infty}\int_{-\infty}^{+\infty}\exp\left[i2\pi\left(xv_x + yv_y + zv_z - vt\right)\right]dv_y\,dv_x$$

(4.18)

follows after substituting equations (4.17) and (4.2) into equation (2.14).

The first term in equation (4.18) describes a superposition of plane waves that propagate in the negative $z$-direction. This possibility is precluded because the $z$-component of the direction of wave propagation is necessarily positive. Consequently

$$C_+D_+ = 0$$

(4.19)

and equation (4.18) reduces to

$$\Psi\left(x,y,z,t\right) = C_+D_-\int_{-\infty}^{+\infty}\int_{-\infty}^{+\infty}\exp\left[i2\pi\left(xv_x + yv_y + zv_z - vt\right)\right]dv_y\,dv_x$$

(4.20)

where

$$C_+D_- \neq 0$$

(4.21)

for propagating waves. Since division by zero is not defined

$$C_+ \neq 0$$

(4.22)

and

$$D_- \neq 0$$

(4.23)

i.e., both factors in equation (4.21) are necessarily non-zero. Furthermore

$$D_+ = 0$$

(4.24)

can be obtained by dividing equation (4.19) by $C_+$. Substitution of equation (4.24) into equation (4.16) yields

$$\tilde{\psi}\left(\nu_x, \nu_y; z\right) = D_- \exp\left(i2\pi z \nu_z\right) \tag{4.25}$$

directly.

The value of $D_-$, given by

$$D_- = \tilde{\psi}\left(\nu_x, \nu_y; 0\right) \tag{4.26}$$

can be obtained by evaluating equation (4.25) on the aperture plane. Furthermore,

$$\tilde{\psi}\left(\nu_x, \nu_y; z\right) = \tilde{\psi}\left(\nu_x, \nu_y; 0\right) \exp\left(i2\pi z \nu_z\right) \tag{4.27}$$

results when equation (4.26) is substituted into equation (4.25). Equation (4.27) describes the angular spectrum that is incident upon the $z$-plane in terms of the angular spectrum on the side of the aperture plane nearest to the observation plane.

After substituting equation (4.27) into equation (4.9)

$$\tilde{P}\left(\nu_x, \nu_y; z\right) = \tilde{\psi}^*\left(\nu_x, \nu_y; 0\right) \tilde{\psi}\left(\nu_x, \nu_y; 0\right) \tag{4.28}$$

can be obtained easily. Equation (4.28) describes the diffraction pattern that forms on the $z$-plane, for arbitrary values of $z$, in terms of the angular spectrum on the side of the aperture plane nearest to the observation plane.

## V. DOUBLE DELTA FUNCTION DECOMPOSITION

### ARBITRARY ORIGIN OF COORDINATES

A definite configuration of quantum objects exists on the side of the aperture plane nearest to the observation plane, i.e., where

$$z = 0 \tag{5.1}$$

As a consequence, equation (4.8) reduces to

$$\psi\left(x, y, 0\right) = \int_{-\infty}^{+\infty} \int_{-\infty}^{+\infty} \tilde{\psi}\left(\nu_x, \nu_y; 0\right) \exp\left[i2\pi\left(x\nu_x + y\nu_y\right)\right] d\nu_y \, d\nu_x \tag{5.2}$$

on the aperture plane. Equation (5.2) is the two-dimensional inverse Fourier transform of $\tilde{\psi}\left(\nu_x, \nu_y; 0\right)$;

$$\tilde{\psi}\left(\nu_x, \nu_y; 0\right) = \int_{-\infty}^{+\infty} \int_{-\infty}^{+\infty} \psi\left(x, y, 0\right) \exp\left[-i2\pi\left(\nu_x x + \nu_y y\right)\right] dy\, dx \tag{5.3}$$

is the corresponding two-dimensional Fourier transform.

Every point in $\psi\left(x, y, 0\right)$ is specified relative to the origin of coordinates in the $\left(x, y, 0\right)$-plane. The origin of coordinates is determined by the arbitrarily located intersection of the $z$-axis with the $\left(x, y, 0\right)$-plane. Undesirable consequences of such arbitrariness can be dealt with by introducing the double delta function decomposition of $\psi\left(x, y, 0\right)$.

**DELTA FUNCTION DECOMPOSITIONS**

Equivalent two-dimensional Dirac delta function decompositions of $\psi\left(x, y, 0\right)$ are given by

$$\psi\left(x, y, 0\right) = \int_{-\infty}^{+\infty} \int_{-\infty}^{+\infty} \psi\left(\alpha, \beta, 0\right) \delta\left(\alpha - x, \beta - y\right) d\beta\, d\alpha \tag{5.4}$$

and

$$\psi\left(x, y, 0\right) = \int_{-\infty}^{+\infty} \int_{-\infty}^{+\infty} \psi\left(\eta, \xi, 0\right) \delta\left(\eta - x, \xi - y\right) d\xi\, d\eta \tag{5.5}$$

where $\left(\alpha, \beta, 0\right)$ and $\left(\eta, \xi, 0\right)$ are arbitrary points within the quantum amplitude $\psi\left(x, y, 0\right)$. Addition of equations (5.4) and (5.5) leads to

$$\psi\left(x, y, 0\right) = \left(\frac{1}{2}\right)\int_{-\infty}^{+\infty} \int_{-\infty}^{+\infty} \int_{-\infty}^{+\infty} \int_{-\infty}^{+\infty} \Big[\psi\left(\alpha, \beta, 0\right) \delta\left(\eta, \xi\right) \delta\left(\alpha - x, \beta - y\right)$$
$$+ \psi\left(\eta, \xi, 0\right) \delta\left(\alpha, \beta\right) \delta\left(\eta - x, \xi - y\right)\Big] d\beta\, d\alpha\, d\xi\, d\eta \tag{5.6}$$

readily. This result expresses the quantum amplitude $\psi\left(x, y, 0\right)$ as a linear combination of two-dimensional Dirac delta function pairs.

**TWO-DIMENSIONAL FOURIER TRANSFORM**

After minor manipulation that includes changing the order of integration and recalling equation (5.3), the two-dimensional Fourier transform of equation (5.6), i.e.,

$$\tilde{\psi}\left(\nu_x, \nu_y; 0\right) = \left(\frac{1}{2}\right) \int_{-\infty}^{+\infty} \int_{-\infty}^{+\infty} \int_{-\infty}^{+\infty} \int_{-\infty}^{+\infty} \int_{-\infty}^{+\infty} \int_{-\infty}^{+\infty} \left[ \psi\left(\alpha, \beta, 0\right) \delta\left(\eta, \xi\right) \delta\left(\alpha - x, \beta - y\right) \right.$$

$$\left. + \psi\left(\eta, \xi, 0\right) \delta\left(\alpha, \beta\right) \delta\left(\eta - x, \xi - y\right) \right] \exp\left[ -i2\pi\left(\nu_x x + \nu_y y\right) \right] dy\, dx\, d\beta\, d\alpha\, d\xi\, d\eta \tag{5.7}$$

can be found. Equation (5.7) reduces to

$$\tilde{\psi}\left(\nu_x, \nu_y; 0\right) = \left(\frac{1}{2}\right) \int_{-\infty}^{+\infty} \int_{-\infty}^{+\infty} \int_{-\infty}^{+\infty} \int_{-\infty}^{+\infty} \left\{ \psi\left(\alpha, \beta, 0\right) \delta\left(\eta, \xi\right) \exp\left[ -i2\pi\left(\nu_x \alpha + \nu_y \beta\right) \right] \right.$$

$$\left. + \psi\left(\eta, \xi, 0\right) \delta\left(\alpha, \beta\right) \exp\left[ -i2\pi\left(\nu_x \eta + \nu_y \xi\right) \right] \right\} d\beta\, d\alpha\, d\xi\, d\eta \tag{5.8}$$

following integration over $x$ and $y$. In turn, equation (5.8) can be written as

$$\tilde{\psi}\left(\nu_x, \nu_y; 0\right) =$$

$$\left(\frac{1}{2}\right) \int_{-\infty}^{+\infty} \int_{-\infty}^{+\infty} \int_{-\infty}^{+\infty} \int_{-\infty}^{+\infty} \left( \psi\left(\alpha, \beta, 0\right) \delta\left(\eta, \xi\right) \exp\left\{ i2\pi\left[ \left(\frac{\eta - \alpha}{2}\right)\nu_x + \left(\frac{\xi - \beta}{2}\right)\nu_y \right] \right\} \right.$$

$$+ \psi\left(\eta, \xi, 0\right) \delta\left(\alpha, \beta\right) \exp\left\{ -i2\pi\left[ \left(\frac{\eta - \alpha}{2}\right)\nu_x + \left(\frac{\xi - \beta}{2}\right)\nu_y \right] \right\} \right)$$

$$X \exp\left\{ -i2\pi\left[ \left(\frac{\eta + \alpha}{2}\right)\nu_x + \left(\frac{\xi + \beta}{2}\right)\nu_y \right] \right\} d\beta\, d\alpha\, d\xi\, d\eta \tag{5.9}$$

after minor manipulation. Each component in the integrand of equation (5.9) describes a two-dimensional complex exponential periodic function multiplied by a two-dimensional complex exponential phase factor.

**APERTURE PLANE QUANTUM AMPLITUDE**

Let $A$ and $B$, where

$$\begin{pmatrix} A \\ B \end{pmatrix} \neq \begin{pmatrix} 0 \\ 0 \end{pmatrix} \tag{5.10}$$

be the quantum amplitudes of the incident quantum objects of temporal frequency $\nu$ at the points $\left(\alpha, \beta, 0\right)$ and $\left(\eta, \xi, 0\right)$, respectively. The corresponding quantum amplitudes on the side of the aperture plane nearest to the observation plane are given by

$$\psi\left(\alpha, \beta, 0\right) = A\delta\left(\alpha - x, \beta - y\right) \tag{5.11}$$

and

$$\psi(\eta,\xi,0) = B\,\delta(\eta-x,\xi-y) \tag{5.12}$$

respectively.

Consider the quantum amplitude associated with the two separated points $(\alpha,\beta,0)$ and $(\eta,\xi,0)$ in a configuration of quantum objects on the side of the aperture plane nearest to the observation plane. This quantum amplitude is given by

$$\psi(x,y,0) = A\,\delta(\alpha-x,\beta-y) + B\,\delta(\eta-x,\xi-y) \tag{5.13}$$

for both points taken together.

The two-dimensional Fourier transform of equation (5.13) is given by

$$\int_{-\infty}^{+\infty}\int_{-\infty}^{+\infty} \psi(x,y,0)\exp\left[-i2\pi(\nu_x x + \nu_y y)\right]dy\,dx = $$
$$\int_{-\infty}^{+\infty}\int_{-\infty}^{+\infty}\left[A\,\delta(\alpha-x,\beta-y) + B\,\delta(\eta-x,\xi-y)\right]\exp\left[-i2\pi(\nu_x x + \nu_y y)\right]dy\,dx \tag{5.14}$$

which yields

$$\tilde{\psi}(\nu_x,\nu_y;0) = A\exp\left[-i2\pi(\nu_x\alpha + \nu_y\beta)\right] + B\exp\left[-i2\pi(\nu_x\eta + \nu_y\xi)\right] \tag{5.15}$$

upon integration. Equation (5.15) can be written as

$$\tilde{\psi}(\nu_x,\nu_y;0) = \exp\left\{-i2\pi\left[\left(\frac{\eta+\alpha}{2}\right)\nu_x + \left(\frac{\xi+\beta}{2}\right)\nu_y\right]\right\}$$
$$X\left(A\exp\left\{i2\pi\left[\left(\frac{\eta-\alpha}{2}\right)\nu_x + \left(\frac{\xi-\beta}{2}\right)\nu_y\right]\right\}\right.$$
$$\left. + B\exp\left\{-i2\pi\left[\left(\frac{\eta-\alpha}{2}\right)\nu_x + \left(\frac{\xi-\beta}{2}\right)\nu_y\right]\right\}\right) \tag{5.16}$$

for easy comparison with equation (5.9).

**QUANTUM AMPLITUDE COMPONENT**

The two-dimensional complex exponential phase factor in equation (5.16) exists because the intersection of the $z$-axis and the $xy$-plane is arbitrary. The distance between the $z$-axis and the midpoint between the points $(\alpha, \beta, 0)$ and $(\eta, \xi, 0)$ is, in accord with the Pythagorean theorem, given by

$$M = \sqrt{\overline{x}^2 + \overline{y}^2} \tag{5.17}$$

where

$$\overline{x} = \frac{\eta + \alpha}{2} \tag{5.18}$$

and

$$\overline{y} = \frac{\xi + \beta}{2} \tag{5.19}$$

are the $x$- and $y$- coordinates, respectively, of the midpoint between the points $(\alpha, \beta, 0)$ and $(\eta, \xi, 0)$. The distance $M$ is not a physical property of the quantum amplitude.

Each two-dimensional complex exponential periodic function in the integrand of equation (5.9) is periodic in two dimensions. One such quantum amplitude component is illustrated in Figure 2. The $x$- and $y$- components of the quantum amplitude spatial period associated with this quantum amplitude are given by

$$T_x = \frac{\eta - \alpha}{2} \tag{5.20}$$

and

$$T_y = \frac{\xi - \beta}{2} \tag{5.21}$$

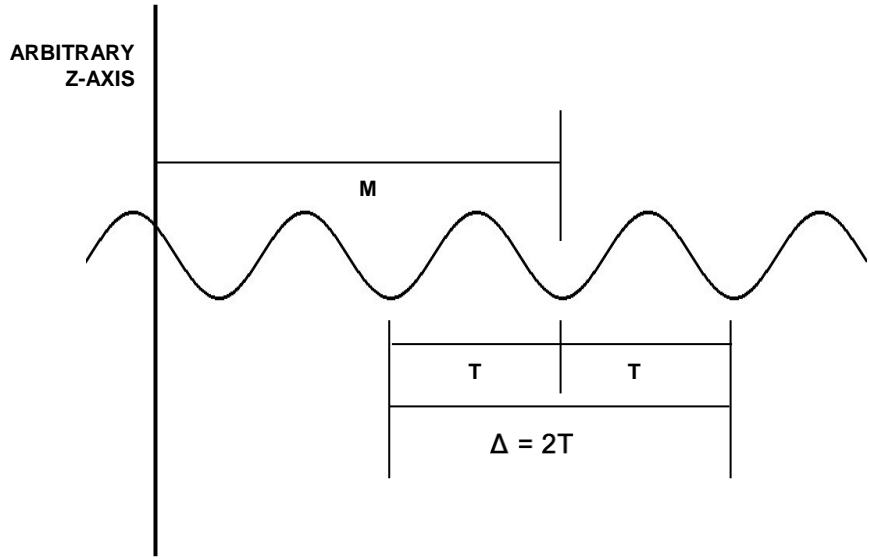

Figure 2. Aperture plane quantum amplitude component.

respectively. The quantum amplitude spatial period is given by

$$T = \sqrt{T_x^{\,2} + T_y^{\,2}} \qquad (5.22)$$

which can be written as

$$T = \sqrt{\left(\frac{\eta - \alpha}{2}\right)^2 + \left(\frac{\xi - \beta}{2}\right)^2} \qquad (5.23)$$

by invoking equations (5.20) and (5.21). The two-dimensional complex exponential periodic functions each represent a physical property of the quantum amplitude.

The distance that separates $(\alpha, \beta, 0)$ and $(\eta, \xi, 0)$ is given by

$$\Delta = \sqrt{\Delta_x^{\,2} + \Delta_y^{\,2}} \qquad (5.24)$$

where

$$\Delta_x = \eta - \alpha \qquad (5.25)$$

and

$$\Delta_y = \xi - \beta \tag{5.26}$$

are the associated distance components in the x- and y- directions, respectively. After substituting equations (5.25) and (5.26) into equation (5.24)

$$\Delta = \sqrt{\left(\eta - \alpha\right)^2 + \left(\xi - \beta\right)^2} \tag{5.27}$$

follows readily. In turn,

$$T = \frac{\Delta}{2} \tag{5.28}$$

can be obtained by substituting equation (5.27) into equation (5.23). Thus, the quantum amplitude spatial period associated with two points in a quantum amplitude is equivalent to one-half the distance between the two points.

The quantum amplitude spatial frequency

$$F = \frac{1}{T} \tag{5.29}$$

can now be defined. After substituting equation (5.28) into equation (5.29)

$$F = \frac{2}{\Delta} \tag{5.30}$$

can be obtained. Thus, the quantum amplitude spatial frequency associated with two points in a quantum distribution is equivalent to twice the reciprocal of the distance between them.

Equation (5.16) reduces to

$$\tilde{\psi}\left(v_x, v_y; 0\right) = \exp\left[-i2\pi\left(\bar{x}v_x + \bar{y}v_y\right)\right] \\ X\left\{A\exp\left[i2\pi\left(T_x v_x + T_y v_y\right)\right] + B\exp\left[-i2\pi\left(T_x v_x + T_y v_y\right)\right]\right\} \tag{5.31}$$

after $\bar{x}$, $\bar{y}$, $T_x$ and $T_y$, given by equations (5.18), (5.19), (5.20), and (5.21), respectively, are recalled. Equation (5.31) can be written as

$$\tilde{\psi}\left(v_x, v_y; 0\right) = \exp\left[-i2\pi\left(\bar{x}v_x + \bar{y}v_y\right)\right] \\ X\left\{\left(A + B\right)\cos\left[2\pi\left(T_x v_x + T_y v_y\right)\right] + i\left(A - B\right)\sin\left[2\pi\left(T_x v_x + T_y v_y\right)\right]\right\} \tag{5.32}$$

by invoking Euler's identity and using trigonometric notation.

Consider illumination by a laterally uniform beam of quantum objects that are transmitted through or reflected from physical features in the aperture plane. For such uniform illumination the quantum amplitudes of the incident quantum objects of temporal frequency $\nu$ at the points $(\alpha, \beta, 0)$ and $(\eta, \xi, 0)$ are equal. Thus

$$A = B \tag{5.33}$$

and equation (5.32) reduces to

$$\tilde{\psi}\left(\nu_x, \nu_y; 0\right) = 2A \exp\left[-i2\pi\left(\bar{x}\nu_x + \bar{y}\nu_y\right)\right] \cos\left[2\pi\left(T_x\nu_x + T_y\nu_y\right)\right] \tag{5.34}$$

Equation (5.34) is applicable even in the extreme case where only one quantum object is involved.

## VI. TWO POINT SEPARATION

After substituting equation (4.27) into equation (4.8)

$$\psi\left(x, y, z\right) = \int_{-\infty}^{+\infty}\int_{-\infty}^{+\infty}\tilde{\psi}\left(\nu_x, \nu_y; 0\right)\exp\left[i2\pi\left(x\nu_x + y\nu_y + z\nu_z\right)\right]d\nu_y\, d\nu_x \tag{6.1}$$

results. Furthermore, substitution of equations (4.1) and (6.1) into equation (2.17) yields

$$\psi\left(x, y, z, t\right) = C_+\int_{-\infty}^{+\infty}\int_{-\infty}^{+\infty}\tilde{\psi}\left(\nu_x, \nu_y; 0\right)\exp\left[i2\pi\left(x\nu_x + y\nu_y + z\nu_z - \nu t\right)\right]d\nu_y\, d\nu_x \tag{6.2}$$

easily. Equation (6.1) describes the quantum amplitude at an arbitrary point between the aperture plane and the observation plane. Equation (6.2) describes a linear combination of plane waves that propagate toward the observation plane.

The spatial frequencies associated with each individual plane wave component in the integrand of equation (6.2) are given by

$$\begin{pmatrix} \nu_x \\ \nu_y \\ \nu_z \end{pmatrix} = \begin{pmatrix} \dfrac{\cos\theta_x}{\lambda} \\ \dfrac{\cos\theta_y}{\lambda} \\ \dfrac{\cos\theta_z}{\lambda} \end{pmatrix} \tag{6.3}$$

a result that follows from equation (2.23) trivially after equation (2.11) has been invoked. Each spatial frequency is associated with a spatial period given by

$$\begin{pmatrix} T_x \\ T_y \\ T_z \end{pmatrix} = \begin{pmatrix} \dfrac{1}{\nu_x} \\ \dfrac{1}{\nu_y} \\ \dfrac{1}{\nu_z} \end{pmatrix} \tag{6.4}$$

by definition. Combination of equations (6.3) and (6.4) leads to

$$\begin{pmatrix} \cos\theta_x \\ \cos\theta_y \\ \cos\theta_z \end{pmatrix} = \begin{pmatrix} \dfrac{\lambda}{T_x} \\ \dfrac{\lambda}{T_y} \\ \dfrac{\lambda}{T_z} \end{pmatrix} \tag{6.5}$$

The propagation direction and a wavefront that is perpendicular to the propagation direction are illustrated in Figure 3 for an individual plane wave component. The wavefront is an infinite plane that is perpendicular to the plane of the figure and that extends out of the plane; only the trace of the wavefront on the plane of the figure is shown in the figure.

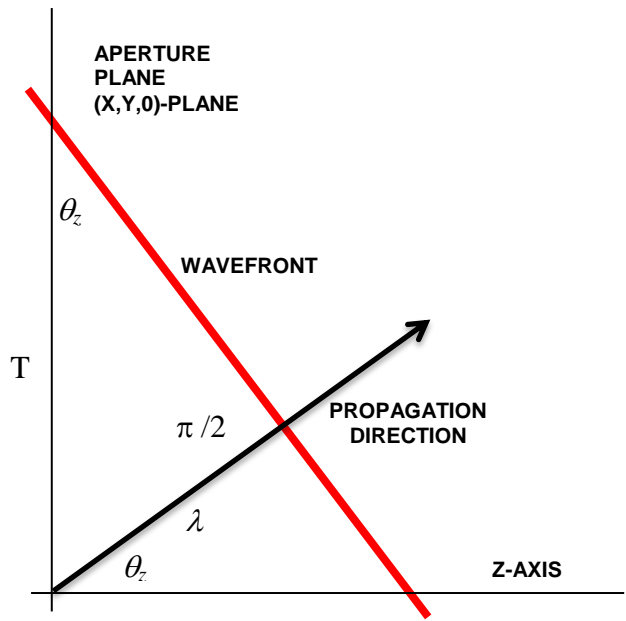

Figure 3. Plane wave propagation geometry.

In Figure 3, as elsewhere in this paper, the $z$-axis is perpendicular to the aperture plane. The aperture plane is rotated about the $z$-axis so that the line defined by the points $(\alpha, \beta, 0)$ and $(\eta, \xi, 0)$ is parallel to the plane of the figure.

The spatial period $T$ is shown as lying along the vertical axis in Figure 3 and is also shown as the hypotenuse of a right triangle. The wavelength $\lambda$ of quantum objects involved is depicted in the figure as the distance between the origin and the wavefront introduced previously. In addition, $\theta_z$ (the propagation angle) is shown as the angle between the direction of wave propagation and the $z$- axis; $\lambda$, $T$ and $\theta_z$ have been introduced previously. The relationship

$$\sin \theta_z = \frac{\lambda}{T} \tag{6.6}$$

can be established by inspecting Figure 3 and applying the definition of an angle's sine. Consequently

$$\theta_z = \sin^{-1}\left(\frac{\lambda}{T}\right) \tag{6.7}$$

is the propagation angle associated with the spatial period $T$. Equation (6.7) can be written as

$$\theta_z = \sin^{-1}\left(\frac{2\lambda}{\Delta}\right) \tag{6.8}$$

where equation (5.28) has been invoked. The propagation angle (in degrees) is plotted in Figure 4 as a function of the separation distance (in wavelengths of the quantum objects involved) between the points $(\alpha, \beta, 0)$ and $(\eta, \xi, 0)$.

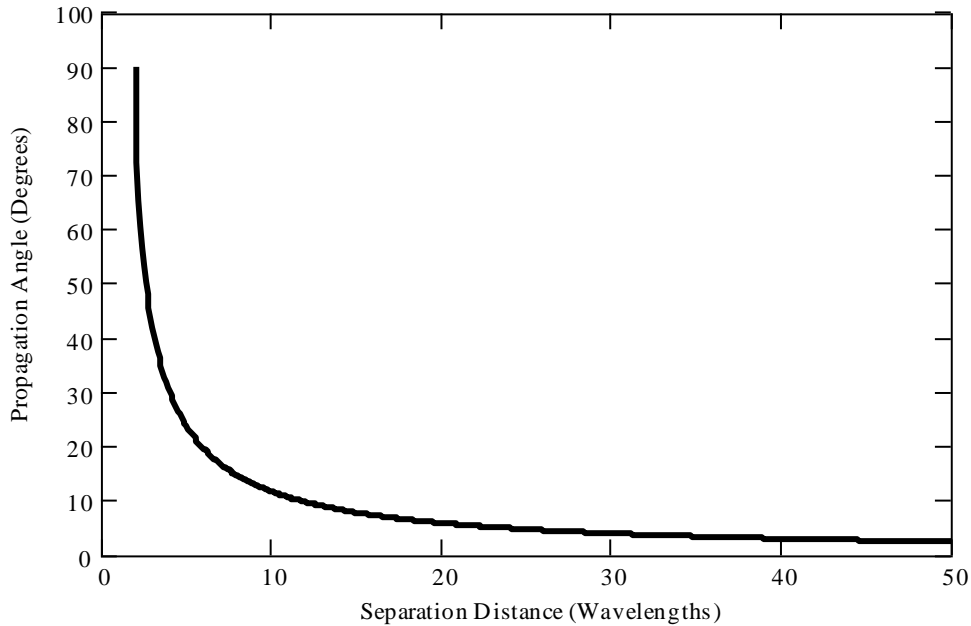

Figure 4. Propagation angle as a function of two-point separation distance.

As shown by equation (2.28) and illustrated in Figure 5, two equi-amplitude plane wave components are associated with each pair of illuminated points in the aperture distribution. For each plane wave component that propagates at an angle $+\theta_z$ relative to the $z$-axis, an equi-amplitude plane wave component propagates at an angle $-\theta_z$ relative to the $z$-axis. Taken together, the two equi-amplitude plane wave components constitute a plane wave pair. The two plane waves superpose to form the angular spectrum described by equation (4.27) at any point $z$ between the aperture plane and the observation plane.

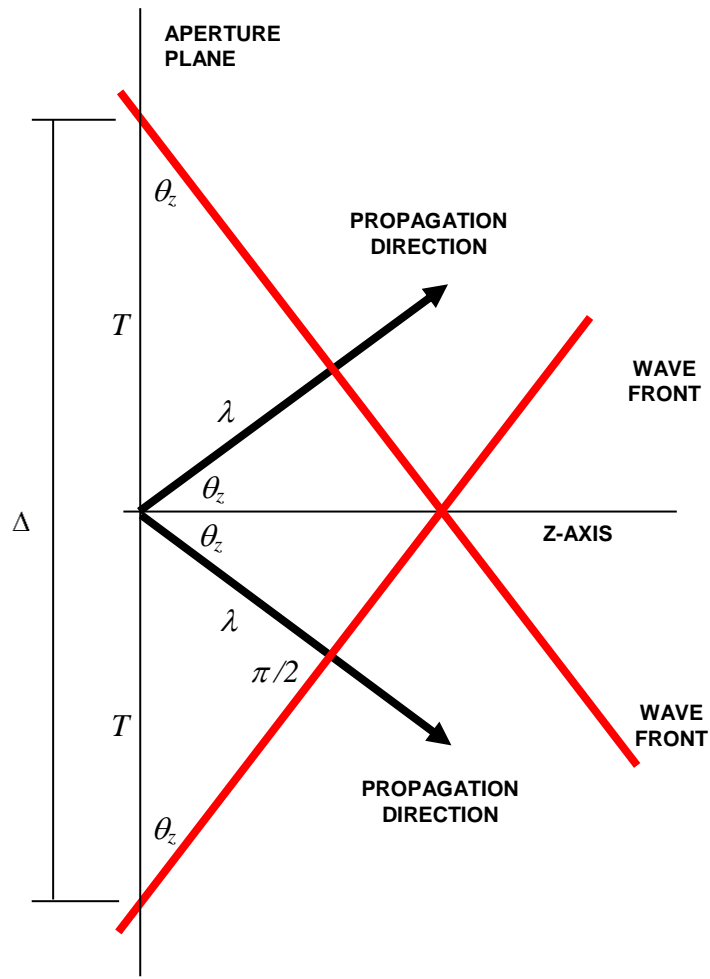

Figure 5. Plane wave pair.

Upon considering equation (6.8) or examining the curve in Figure 4 it can be determined readily that two illuminated points in the aperture plane are required for quantum objects to propagate away from the aperture plane. In addition, the minimum separation distance between the two illuminated points exceeds two wavelengths of the quantum objects involved. Consequently, propagation of quantum objects away from a single point (as hypothesized by the Huygens-Fresnel principle) does not occur.

As indicated previously, diffraction patterns are formed by the apparently random arrival of one quantum object at a time on the observation screen. Thus, individual quantum objects are spread out to cover both required illuminated points in the aperture plane.

## VII. CONCLUSION

A unified theory of wave-particle duality, a derivation of the time independent and time dependent Schrödinger equations, and a subsequent treatment of quantum diffraction have been presented in this paper.

Quantum objects are inseparably associated with wave-like properties and particle-like properties; this twofold character is known as wave-particle duality. The name (electron, photon, atom, molecule, etc.) linked to its particle-like properties is used to identify a quantum object. Classically, wave-like properties are associated with spatial extension while particle-like properties are associated with a point-like locality. As shown in this paper, the wave-like properties and the particle-like properties linked to a quantum object are both associated with spatial extension. Quantum objects are nonlocal.

After developing the unified theory of wave-particle duality, straightforward derivations that led unequivocally to the Schrödinger equations were presented. Thus, the time-independent and time-dependent Schrödinger equations were derived from fundamental physics in this paper. Reliance on plausibility arguments to justify the Schrödinger equations is unwarranted.

In accord with the unified theory of wave-particle duality, quantum objects are linked to particle-like properties, but not to particles. The Schrödinger equations are field equations, not particle equations. Rather than describing particle motion, the time-dependent Schrödinger equation describes a time-dependent field $\Psi(\mathbf{r}, t)$ throughout a spatial region. The Schrödinger field is a space-filling physical field whose value at any spatial point is the probability amplitude for an interaction to occur within an infinitesimal region surrounding the point.

The Schrödinger equations provide a basis for analyzing many kinds of systems (molecular, atomic, and nuclear) in a particular inertial reference frame. The success of the Schrödinger equations constitutes a basis for accepting them, their derivations, and the unified theory of wave-particle duality which makes such derivations possible. This acceptance is completely justified in favored inertial reference frames.

In addition to being justified in a particular inertial reference frame, the Schrödinger equations are obtained by reduction from a Lorentz invariant equation – the optical differential wave equation – in this paper. Consequently, the Schrödinger equations are relativistically invariant and constitute laws of physics.

As a consequence of their wave-like properties, quantum objects can exhibit diffraction phenomena. Quantum diffraction occurs when the lateral extent of propagating waves linked to one or more quantum objects is spatially restricted. When waves pass through a region where they are spatially restricted they spread into regions that are not directly exposed to them. This phenomenon is called diffraction and is a property of all kinds of propagating waves.

During the diffraction process treated in this paper, quantum objects that are incident upon an aperture plane are transmitted through or reflected from physical features in the aperture plane. The quantum objects are consequently distributed in a definite configuration on the side of the aperture plane nearest to the observation plane. Subsequently, various quantum objects propagate from the aperture plane to the observation plane where they contribute to diffraction pattern formation.

At least two illuminated points in the aperture plane are required to support propagation of quantum objects away from the aperture plane toward the observation plane. The minimum separation distance between any two such points exceeds two wavelengths of the quantum object involved. Propagation of quantum objects away from a single point in the aperture plane does not occur.

Quantum diffraction occurs for any number of incident quantum objects, even when that number is reduced to one. Quantum objects are not geometrical point-like localized objects; rather, they are spread-out and nonlocal.

The quantum amplitude components linked to a quantum object define the relative interaction probability density for the interaction of the quantum object at a particular location on the observation surface. Each quantum object is associated with the entire relative interaction probability density distribution. The diffraction pattern that is formed is defined by the relative interaction probability density.

Diffraction patterns are formed by the apparently random detection of individual quantum objects. A diffraction pattern that is formed in a sufficiently feeble manner shows no evidence of having wave-like properties. Rather, diffraction patterns are built up by individual quantum objects that act independently. An inherent granularity exists during the early portion of the diffraction pattern build-up.

A very large number of quantum objects contribute when diffraction pattern formation is complete. The inherent granularity linked to the individual quantum objects vanishes when the diffraction pattern is completely formed.

A diffraction pattern that is formed in a sufficiently strong manner shows no evidence of being built up by individual quantum objects. Rather, the entire diffraction pattern appears to be formed as a single occurrence.

## APPENDIX A: WAVE-PARTICLE DUALITY PARADOX

Quantum diffraction experiments[18] provide well-known demonstrations of the inseparability of the wave-like and particle-like properties of quantum objects. These demonstrations can be described easily, but understanding them has awaited the theory introduced in this paper.

The wave nature of quantum objects is demonstrated by the diffraction pattern that forms on an observation screen. However, the quantum objects that contribute to diffraction pattern formation are always detected as discrete entities. Individual quantum objects that act independently of each other accumulate to build up the diffraction pattern.

### NAIRZ, ARNDT AND ZEILINGER

Early development of the wave-particle duality concept has been recently reviewed[19] by Nairz, Arndt and Zeilinger as part of an investigation relevant to the limit of the concept's applicability. Salient quotations from this study follow:

> "At the beginning of the 20th century several important discoveries were made leading to a set of mind-boggling questions and experiments that seemed to escape any answers based on classical, pre-quantum physics. The first were the discoveries[20, 21, 22] that implied that optical radiation has to be composed of discrete energy packages that can be well localized in space and time. This localization was in marked contrast to the existing knowledge based on Maxwell's theory which successfully represented light as electromagnetic waves. The second and complementary breakthrough was the theoretical result by deBroglie[23] and the experimental demonstration by Davisson and Germer[24] that massive particles also propagate in a wavelike manner.

> "Both statements were stunning at the time they were proposed and both keep us thinking even today because we generally associate the notion of point-like locality with a particle while we attribute spatial extension to a wave."

Later, Nairz, Arndt and Zeilinger go on to state:

> "Based on these historical achievements we ask how far we might be able to extend such quantum experiments and for what kind of objects we might be able to show the wave-particle duality. Recently, a new set of experiments exceeding the mass and complexity of the previously used objects by about an order of magnitude has been developed in our laboratory. These experiments with the fullerene molecule $C_{60}$ will be described in Sec. II."

Still Later, the authors conclude:

"Quantum phenomena become increasingly important and the limit to which we may be able to confirm all quantum principles experimentally is still an open question. The discussion of our fullerene experiments lets us demonstrate the basic wave-particle duality for the most massive, most complex, and most 'classical' single object so far."

**FEYNMAN**

Before the fullerene work was reported, Feynman[25] summarized the situation thus:

"Things on a very small scale behave like nothing you have any direct experience about. They do not behave like waves, they do not behave like particles, they do not behave like clouds, or billiard balls, or weights on springs, or like anything you have ever seen.

"Newton thought that light was made up of particles, but then it was discovered that it behaves like a wave. Later, however (in the beginning of the twentieth century) it was found that light did indeed sometimes behave like a particle. Historically, the electron, for example, was thought to behave like a particle, and then it was found in many respects it behaved like a wave. So it really behaves like neither. Now we have given up. We say: 'It is like *neither*.'

"There is one lucky break, however – electrons behave just like light. The quantum behavior of atomic objects (electrons, protons, neutrons, photons, and so on) is the same for all, they are all 'particle waves' or whatever you want to call them. So what we learn about the properties of electrons (which we shall use for our examples) will apply also to all 'particles' including photons of light."

**BOHR**

The principle of complementarity, introduced by Niels Bohr, is commonly invoked in quantum mechanics to resolve the wave-particle duality paradox. Albert Messiah has phrased this principle in the following way[26]:

"The description of the physical properties of microscopic objects in classical language requires pairs of complementary variables; the accuracy in one member of the pair cannot be improved without a corresponding loss in the accuracy of the other member."

A little later, Messiah resolves the wave-particle duality paradox (Messiah uses the word *corpuscle* rather than the word *particle*) in the following way[27]:

> "If one adopts the principle of complementarity, the wave-corpuscle duality ceases to be paradoxical; the wave aspect and the corpuscular aspect are two complementary aspects which are exhibited only in mutually exclusive experimental arrangements. Any attempt to reveal one of the two aspects requires a modification of the experimental set-up which destroys any possibility of observing the other aspect."

This view seems to be widely accepted by present-day physicists.

**RABINOWITZ**

Mario Rabinowitz has examined[28] experimental and theoretical attempts to test the prediction that any detector capable of determining the path taken by a particle through one slit or the other of a two-slit aperture will destroy the diffraction pattern. Quoting Rabinowitz:

> "The wave-particle duality is the main point of demarcation between quantum and classical physics, and is the quintessential mystery of quantum mechanics. Young's two-slit diffraction experiment is the arch prototype of actual and gedanken experiments used as a testing ground of this duality. Quantum mechanics predicts that any detector capable of determining the path taken by a particle through one or the other of a two-slit plate will destroy the diffraction pattern. We will examine both the experimental and theoretical attempts to test this assertion, including a new kind of experiment, and to grasp the underlying truth behind this mystery from the earliest days to the present. Where positions differ, the views of both sides are presented in a balanced approach."

Rabinowitz leaves the wave-particle duality paradox unresolved.

## APPENDIX B: JUSTIFICATION OF THE SCHRÖDINGER EQUATIONS

No fundamental derivation of the Schrödinger equations existed before the unified theory of wave-particle duality was developed. Rather, subjective plausibility arguments that vary from author to author are ordinarily used to justify the Schrödinger equations. The Schrödinger equations provide a basis for successfully analyzing many kinds of systems (molecular, atomic, and nuclear). This success constitutes the basis for widespread pragmatic acceptance of them. Some authors provide justifications for the acceptance of the Schrödinger equations that may be of some interest.

**FEYNMAN**

Richard Feynman commented concerning the time-dependent Schrödinger equation[29] in the following manner:

"We do not intend to have you think we have derived the [time-dependent] Schrödinger equation but only wish to show you one way of thinking about it. When Schrödinger first wrote it down, he gave a kind of derivation based on some heuristic arguments and some brilliant intuitive guesses. Some of the arguments he used were even false, but that does not matter; the only important thing is that the ultimate equation gives you a correct description of nature."

A little later[30], Feynman proffered the following observation relevant to the time-dependent Schrödinger equation:

"Where did we get that from? Nowhere. It's not possible to derive it from anything you know. It came out of the mind of Schrödinger, invented in his struggle to find an understanding of the experimental observations of the real world."

**FRENCH AND TAYLOR**

At the end of a very good effort to introduce the Schrödinger equations[31] A. P. French and Edwin F. Taylor comment as follows:

"Clearly we have not been inexorably driven to [the Schrödinger equations] any more than Schrödinger was in his argument from analogy. One can construct many other differential equations that embody the dynamical relations expressed [by the equations used to make the Schrödinger equations appear plausible]. But the Schrödinger equations, besides being the mathematically simplest equations that satisfy the requirements, have other properties that cause them to be preferred over all other possibilities:

1. They have the property of *linearity*, so if $\Psi_1$ and $\Psi_2$ are specific solutions to one of these equations, then any linear combination of them is a solution of the same equation. This property of *superposition* is one of the most basic properties of waves.

2. Their solutions are (as we shall see [later]) perfectly suited to the interpretation of $\Psi$ as a probability amplitude.

"Finally, there is all the accumulated evidence that the Schrödinger equations *work*; they provide the basis for a correct analysis of all kinds of molecular, atomic, and nuclear systems. Whatever questionable features there may be in the manner of their formulation are swept away in the evidence of their manifest success."

The presentation by French and Taylor may be the most lucid justification (not derivation) of the Schrödinger equations that is available.

**MERZBACHER**

Merzbacher offers the following justification[32] for adopting the Schrödinger equations:

> "Equipped with all the hindsight which the reading of this history provides, we can try to justify the dynamical law of quantum mechanics without following the historical development. Of course, a certain amount of guessing is inevitable in obtaining as general a law as we are hoping for, and it would be misleading to obscure this fact by pretending that the basic equations can be *derived*. The best that can be done here is to make the final dynamical law, the wave equation of quantum mechanics, appear reasonable."

Merzbacher used linear independence[33] as an ingredient of his presentation:

> "··· the linear independence insures that any harmonic plane wave can be written at all times as a linear combination of $\Psi_1$ and $\Psi_2$. It is then eminently reasonable to demand that the motion of a free particle moving in the positive *x*-direction should at all times be represented by a pure plane wave of the type $\Psi_1$ without any admixture of $\Psi_2$, which moves in the opposite direction."

**VARIETY OF JUSTIFICATIONS**

Many justifications for the Schrödinger equations exist in addition to those which have been considered. Treatment of these additional justifications is beyond the scope of this appendix.